\documentclass[12pt,letterpaper]{article}
\usepackage[usenames, dvipsnames]{xcolor}
\usepackage{jcapmod}

\usepackage{tocloft}
\usepackage[]{todonotes}
\usepackage{verbatim}
\usepackage{mathrsfs}
\usepackage{cleveref}
\usepackage[shortlabels]{enumitem}

\usepackage{pgf}
\usepackage{tikz-cd}
\usetikzlibrary{shapes,arrows}
\usetikzlibrary{calc}
\usepackage{pgfplots}
\pgfplotsset{compat=1.12}
\usepackage{tikz-3dplot}

\usepackage{tikz}
\usepackage{setspace,caption}
\usepackage{lipsum}
\usetikzlibrary{matrix,arrows,calc}
\makeatletter
\newsavebox\myboxA
\newsavebox\myboxB
\newlength\mylenA

\definecolor{cornellRed}{HTML}{B31B1B}
\definecolor{cornellBlue}{HTML}{0068AC}
\definecolor{cornellGreen}{HTML}{6EB43F}

\usepackage{bbm} 					
\usepackage{slashed} 				
\usepackage{graphicx}				
\usepackage{subcaption}			
\usepackage{psfrag}				
\usepackage{tensor}				
\usepackage{fouridx}				
\usepackage{bm}					
\usepackage{mdframed}				
\usepackage{multirow}				
\usepackage{soul}					
\usepackage{bbold}				
\usepackage{multicol}				
\usepackage{tikz-cd}
\usepackage{rotating}

\usetikzlibrary{arrows}

\tikzset{
commutative diagrams/.cd,
arrow style=tikz,
diagrams={>=latex}}

\usepackage{amsmath}
\usepackage{amssymb}
\usepackage{amsfonts}
\usepackage{mathtools}

\usepackage{feynmf}
\usepackage{marvosym}

\usepackage{import}

\newcommand*\xoverline[2][0.75]{%
    \sbox{\myboxA}{$\m@th#2$}%
    \setbox\myboxB\null
    \ht\myboxB=\ht\myboxA%
    \dp\myboxB=\dp\myboxA%
    \wd\myboxB=#1\wd\myboxA
    \sbox\myboxB{$\m@th\overline{\copy\myboxB}$}
    \setlength\mylenA{\the\wd\myboxA}
    \addtolength\mylenA{-\the\wd\myboxB}%
    \ifdim\wd\myboxB<\wd\myboxA%
       \rlap{\hskip 0.5\mylenA\usebox\myboxB}{\usebox\myboxA}%
    \else
        \hskip -0.5\mylenA\rlap{\usebox\myboxA}{\hskip 0.5\mylenA\usebox\myboxB}%
    \fi}
\makeatother

\newcommand{\im}{\mathrm{Im}\,}  
\newcommand{\re}{\,\mathrm{Re}\,}

\newcommand{\tr}{\,\mathrm{tr}}

\definecolor{cobalt}{RGB}{44, 98, 120}
\definecolor{celadon}{rgb}{0.67, 0.88, 0.69}
\definecolor{dm}{cmyk}{.20, 0, .30, 0}
\definecolor{burgundy}{rgb}{0.5, 0.0, 0.13}
\definecolor{plotBlue}{RGB}{94, 130, 181}

\hypersetup{
  colorlinks,
  citecolor=Violet,
  linkcolor=cobalt,
  urlcolor=Blue}

\DeclareSymbolFontAlphabet{\mathbb}{AMSb}

\newif\iffastcompile

\fastcompilefalse

\iffastcompile
\newcommand{\mk}[1]{}
\newcommand{\lm}[1]{}
\else
\newcommand{\mk}[1]{\todo[color=burgundy!30, size=\scriptsize, bordercolor=burgundy!30]{MK: #1}}
\newcommand{\lm}[1]{\todo[color=dm!90, size=\scriptsize, bordercolor=dm!90]{LM: #1}}

\fi

\newcommand{\email}[1]{\href{mailto:#1}{#1}}

\ProvideTextCommandDefault{\Dbar}{%
\leavevmode\lower.5ex\rlap{\hskip-.07em\accent"16}D%
}

\begin{document}
	\newcommand{\main}{.}
\begin{titlepage}

\setcounter{page}{1} \baselineskip=15.5pt \thispagestyle{empty}
\setcounter{tocdepth}{2}

\bigskip\

\vspace{1cm}
\begin{center}
{\fontsize{22}{28} \bfseries Monodromy Charge in D7-brane Inflation}
\end{center}

\vspace{0.45cm}

\begin{center}
\scalebox{0.95}[0.95]{{\fontsize{14}{30}\selectfont Manki Kim and Liam McAllister \vspace{0.25cm}}}

\end{center}

\begin{center}

\textsl{Department of Physics, Cornell University, Ithaca, NY 14853, USA}\\

\vspace{0.25cm}

\vskip .3cm
\email{\tt mk2427@cornell.edu, mcallister@cornell.edu }
\end{center}

\vspace{0.6cm}
\noindent
In axion monodromy inflation, traversing $N$ axion periods corresponds to discharging $N$ units of a quantized charge.  In certain models with moving D7-branes, such as Higgs-otic inflation, this monodromy charge is D3-brane charge induced on the D7-branes.  The stress-energy of the induced charge affects the internal space, changing the inflaton potential and potentially limiting the field range.
We compute the backreaction of induced D3-brane charge in Higgs-otic inflation.  The effect on the nonperturbative superpotential is dramatic even for $N=1$, and may preclude large-field inflation in this model in the absence of a mechanism to control the backreaction.

\noindent
\vspace{2.1cm}

\noindent\today

\end{titlepage}
\tableofcontents\newpage

\section{Introduction}

Inflationary models involving super-Planckian displacements provide a striking connection between quantum gravity and observable phenomena.  Upper limits on primordial B-mode polarization in the CMB have excluded some models of large-field inflation, but others remain viable \cite{Akrami:2018odb}.  At the same time, the theoretical question of the status of super-Planckian displacements in quantum gravity remains unresolved, despite much activity.

Large-field inflation is readily described in effective field theory, but crucially relies on assumptions about symmetries in quantum gravity.  A prototypical example is the shift symmetry of an axion with decay constant $f \gg M_{\rm{pl}}$ \cite{FFO}.
No assumption about quantum gravity that is sufficient to protect large-field inflation has yet been put on indisputably solid footing in string theory: on the contrary, general expectations about the destruction of global symmetry charges by black holes, as well as conjectures about Weak Gravity and about moduli spaces in quantum gravity \cite{Ooguri:2006in,ArkaniHamed:2006dz},
suggest that controlling a super-Planckian displacement in a quantum gravity theory is difficult.  In view of these results, ignoring the problem of ultraviolet completion and studying large-field inflation solely from the bottom up appears untenable.

A practical way forward is to search for candidate realizations of large-field inflation in compactifications of string theory, and to investigate their characteristics and limitations.  To shed light on the question of interest, these realizations should be sufficiently explicit, and sufficiently well-controlled, so that quantum gravity corrections to the inflaton action can be computed.

In this work we study models of large-field inflation in string theory in which the inflaton is the position of a D7-brane.  We focus on
D7-brane monodromy scenarios, such as Higgs-otic inflation \cite{Ibanez:2014swa}, in which the D7-brane repeatedly traverses a loop in the internal space, discharging an induced charge or flux, and reducing the four-dimensional energy density, with each cycle.  Compared to other scenarios for axion monodromy inflation in string theory, an advantage of existing D7-brane models is that the compactification can be a simple and comparatively explicit toroidal orientifold.
In this setting, one can carefully examine effects that might interfere with achieving a super-Planckian displacement.

Arguably the most dangerous effect in axion monodromy inflation is \emph{backreaction of monodromy charge}.
Transporting the inflaton field $N$ times around a loop in configuration space leads to the accumulation of $N$ units of physical, quantized charge, corresponding for example to D-brane charge carried by branes or fluxes.  This monodromy charge is the order parameter measuring displacement from the minimum of the inflaton potential.
The stress-energy of the monodromy charge is a leading source in the four-dimensional Einstein equations, and in a successful model this stress-energy drives inflationary expansion.  At the same time, the monodromy charge is a source for the Einstein equations in the internal six dimensions.  We refer to the resulting effects on the internal space as `backreaction of monodromy charge', and we use the term `probe approximation' to describe the approach of neglecting the backreaction.

One of our main
conclusions
is that in D7-brane axion monodromy inflation, the probe approximation is \emph{not} a valid or consistent approximation.
The problem of backreaction of monodromy charge was already emphasized in \cite{MSW} and its implications were the main subject of \cite{FMPWX,MSSSW}, but because these works examined axion monodromy on NS5-branes \cite{MSW} --- a scenario requiring a rather complicated warped throat compactification --- some have suggested that backreaction of monodromy charge may be a particular defect of the NS5-brane model, and may be negligible in all F-term axion monodromy models \cite{Marchesano:2014mla}.
Our analysis excludes this possibility.  We find that the backreaction of monodromy charge is, if anything, even more visible and more dangerous in D7-brane monodromy on toroidal orientifolds than it is in the NS5-brane case: it was shown in \cite{FMPWX,MSSSW} that by fine-tuning the position of an NS5-brane pair in a warped throat, the leading backreaction effects can be mitigated, but there is no obvious analogue of this mechanism in a toroidal orientifold.  We do not rule out the possible existence of a mechanism for ameliorating backreaction in D7-brane inflation, but in our view, inventing and establishing such a mechanism is a prerequisite to any claim of large-field inflation in this setting.  On the other hand, although our work naturally generalizes to other models with monodromy charge localized on D-branes or NS-branes, backreaction may be less problematic in scenarios with delocalized monodromy charge, e.g.~in the form of bulk fluxes \cite{McAllister:2014mpa}.\footnote{We thank E.~Silverstein for emphasizing this point.}

The ten-dimensional backreaction we consider here should be carefully distinguished from the four-dimensional backreaction studied in
\cite{Baume:2016psm,Valenzuela:2016yny,Blumenhagen:2017cxt}, which involves non-linear interactions among moduli fields in four-dimensional theories, e.g.~shifts of saxion vevs following large axion displacements, along the lines of \cite{flattening}.  We are examining the effects of localized sources in the ten-dimensional equations of motion: these lead to couplings that are difficult or impossible to compute in the four-dimensional theory obtained by dimensional reduction in the probe approximation.
In particular, ten-dimensional backreaction effects are not readily computed in a Kaloper-Sorbo \cite{Kaloper:2008fb} description of axion monodromy inflation in a four-dimensional effective theory, and should be understood instead as \emph{ultraviolet inputs}
to such a theory.
In particular, a primary aim of the present work is to compute, in ten-dimensional supergravity, the precise form of the Pfaffian prefactors \eqref{eqn:the one loop pfaffian} that were approximated by constants in \cite{Baume:2016psm,Valenzuela:2016yny,Blumenhagen:2017cxt} and were modeled phenomenologically in \cite{Ruehle:2017one}.  Our results \eqref{eqn:the one loop pfaffian}, \eqref{eqn:the one loop pfaffian explicit} can then be taken as inputs for analyses in the frameworks of \cite{Baume:2016psm,Valenzuela:2016yny,Blumenhagen:2017cxt,Ruehle:2017one}.

The organization of this note is as follows.  In \S\ref{sec:modelbackground} we review the construction of Higgs-otic inflation \cite{Ibanez:2014swa}.  In \S\ref{sec:backreaction} we compute the backreaction of induced D3-brane charge in configurations of moving D7-branes.
We describe the impact of this effect on Higgs-otic inflation in \S\ref{sec:implications}, and we also comment on a related issue in fluxbrane inflation.  Our conclusions appear in \S\ref{sec:conclusions}.
Appendix \ref{app:convention} gives our conventions for differential forms, and Appendix \ref{app:green} collects a few results about Green's functions in toroidal orientifolds.

\section{Higgs-otic Inflation} \label{sec:modelbackground}

We begin by recalling key elements of the Higgs-otic inflation scenario \cite{Ibanez:2014swa,Bielleman:2015lka,Bielleman:2016olv}.  For the phenomenology of these models, which we will not review, we refer the interested reader to the original references \cite{Ibanez:2014swa,Bielleman:2015lka,Bielleman:2016olv}.  Related constructions include \cite{Dasgupta:2002ew,Dasgupta:2004dw,Haack:2008yb,Marchesano:2014mla,Hebecker:2014eua}.

Higgs-otic inflation
is a construction of chaotic inflation in type IIB string theory via monodromy. The inflaton field is identified as the position of a D7-brane wrapping a four-cycle in a flux compactification.
As the D7-brane moves through a background of three-form flux, it accumulates induced anti-D3-brane charge, breaking supersymmetry and creating a potential.  The idea is to choose the geometry and flux in such a way that the D7-brane can repeatedly travel around a one-cycle in the compactification, acquiring more induced anti-D3-brane charge with each cycle.  In other words, the D7-brane couplings to the background flux introduce monodromy, and the order parameter for the monodromy is the amount
$Q^{\overline{D3}}$ of induced anti-D3-brane charge on the D7-brane.

\subsection{Setup}

We will examine Higgs-otic inflation in the context of compactifications of type IIB string theory on toroidal orientifolds.
In the conventions of \cite{Polchinski:2000uf}, the type IIB supergravity action in Einstein frame takes the manifestly $SL(2,\Bbb{Z})$-invariant form
\begin{align} \label{sugraaction}
S_{IIB}=&\frac{1}{2\kappa_{10}^2}\int_{\Bbb{R}^{1,3}\times X} \star_{10}\mathcal{R}-\frac{1}{2(\im\tau)^2}d\tau\wedge\star_{10}d\bar{\tau}-\frac{1}{2\im\tau}G_3\wedge\star_{10}\bar{G}_3-\frac{1}{4}\tilde{F}_5\wedge\star_{10}\tilde{F}_5\nonumber\\
&+\frac{1}{8i\kappa_{10}^2}\int_{\Bbb{R}^{1,3}\times X}\frac{1}{\im\tau}C_4\wedge G_3\wedge \bar{G}_3+S_{loc}\,.
\end{align}
We consider an ansatz for the metric and Ramond-Ramond five-form of the form
\begin{align}
ds^2=&h^{-1/2}(z)ds^2_{\Bbb{R}^{1,3}}+h^{1/2}(z)ds^2_{X},\label{eqn:metric ansatz}\\
\tilde{F}_5=&(1+\star_{10})d\alpha(z)\wedge\sqrt{-\det(g)}dx^0\wedge dx^1\wedge dx^2\wedge dx^3\nonumber,
\end{align}
where $z$ denotes the coordinates on the internal space $X$.
We denote the Hodge star operators in ten dimensions, on $X$, and on a divisor $D \subset X$ by $\star_{10}$, $\star_{6}$, and $\star_{4}$, respectively.
We also define
\begin{equation}
G_{\pm}=\frac{(\star_6\pm i)}{2}G_3,
\end{equation} and refer to $G_+$ and $G_-$ as imaginary self-dual (ISD) and imaginary anti-self-dual (IASD) flux, respectively.   See Appendix \ref{app:convention} for more details of our conventions.

In \cite{Ibanez:2014swa} $G$ was assumed to be a constant  ISD  flux, while \cite{Bielleman:2016olv} generalized $G$ to a linear combination of ISD and  IASD  fluxes.  For simplicity, in this section we consider an ISD background with $G_-=0$, $h^{-1}=\alpha$, and constant axio-dilaton field $\tau$; our main analysis in \S\ref{sec:backreaction} is robust to relaxing these restrictions.

\subsection{Magnetized D-brane action} \label{app:dbraneaction}

Consider a D7-brane that fills the noncompact spacetime and wraps a divisor $D \subset X$.
A general two-form flux $\mathcal{F}$ on the D7-brane can be written as the sum of self-dual (SD) and anti-self-dual (ASD) components:
\begin{equation}\label{eqn:general 2 flux}
\mathcal{F}=(1+\star_4)\mathcal{F}/2+(1-\star_4)\mathcal{F}/2=\mathcal{F}_++\mathcal{F}_-.
\end{equation}
We will refer to a D7-brane carrying nontrivial worldvolume flux $\mathcal{F}$ as being \emph{magnetized}.
In this section we examine the Dirac-Born-Infeld (DBI) and Chern-Simons (CS) actions of a magnetized D7-brane.

Viewing the two-form flux on $D$ as a $4\times4$ skew-symmetric matrix, and writing the metric on $D$ as $g$, we have the identities
\begin{align}
\det(I+g^{-1} \mathcal{F})=&1-\frac{1}{2} \tr (g^{-1}\mathcal{F})^2+\det(g^{-1}\mathcal{F}),
\end{align}
\begin{equation}
-\frac{1}{2}\int_D\text{Vol}_D \tr (g^{-1}\mathcal{F})^2=\int_D \mathcal{F}\wedge \star_4\mathcal{F}.
\end{equation}
It follows that
\begin{align}
\det(I+g^{-1}\mathcal{F})^{1/2}=&1-\frac{1}{4}\tr (g^{-1}\mathcal{F})^2+\frac{1}{2}\det(g^{-1}\mathcal{F})-\frac{1}{32}\left[ \tr (g^{-1}\mathcal{F})^2\right]^2+\mathcal{O}(\mathcal{F}^6).
\end{align}
Note that the above expansion is exact up to $\mathcal{O}(\mathcal{F}^2)$ if $\mathcal{F}=\pm\star_4\mathcal{F}.$

We can now expand the DBI+CS actions of a static D7-brane in an ISD background, written in Einstein frame, up to $\mathcal{O}(\mathcal{F}^2)$:
\begin{align}
S_{D7}=&-\mu_7\int_{\Bbb{R}^{1,3}\times D}\text{Vol}_{\Bbb{R}^{1,3}}\wedge\text{Vol}_D (\im\tau)^{-1}\det\left(I+(\im\tau)^{1/2}g^{-1}\mathcal{F}\right)^{1/2}\nonumber\\
&+\mu_7\int_{\Bbb{R}^{1,3}\times D}C_8+C_6\wedge\mathcal{F}+\frac{1}{2}C_4\wedge\mathcal{F}\wedge\mathcal{F}\\
=&-\mu_7\int_{\Bbb{R}^{1,3}\times D}\text{Vol}_{\Bbb{R}^{1,3}}\wedge\frac{1}{2} \Bigl((\im\tau)^{-1}\mathcal{J}\wedge \mathcal{J}+\mathcal{F}\wedge\star_4\mathcal{F}\Bigr)\nonumber\\
&+\mu_7\int_{\Bbb{R}^{1,3}\times D}C_8+\frac{1}{2}C_4\wedge\mathcal{F}\wedge\mathcal{F}+\mathcal{O}(\mathcal{F}^4)\,. \label{expandedact}
\end{align}
Here $\text{Vol}_{\Bbb{R}^{1,3}}$ is the volume in the metric $h^{-1/2}g_{\mu\nu}$,
and similarly the Hermitian form\footnote{The Hermitian form $\mathcal{J}$ is a K\"ahler form if $d\mathcal{J}=0.$} $\mathcal{J}$ corresponds to the full internal metric including the warp factor, and obeys $\frac{1}{2}\mathcal{J}\wedge \mathcal{J}=\text{Vol}_D.$
We have dropped the $C_6\wedge\mathcal{F}$\ term because $C_6$ can be fixed to be zero in an ISD background.
From the Chern-Simons term involving $C_4$ in \eqref{expandedact} it is clear that an SD flux on a D7-brane induces D3-brane
charge, whereas an ASD flux induces $\overline{D3}$-brane charge.

The candidate inflaton potential arises from the terms in the D7-brane action \eqref{expandedact} that are quadratic in $\mathcal{F}$:
\begin{align}
S_{\mathcal{F}^2}=&-\mu_7 \int_{\Bbb{R}^{1,3}\times D}\left(\text{Vol}_{\Bbb{R}^{1,3}}-C_4\right)\wedge \frac{1}{2}\mathcal{F}_+\wedge \star_4\mathcal{F}_+-\mu_7 \int_{\Bbb{R}^{1,3}\times D}\left(\text{Vol}_{\Bbb{R}^{1,3}}+C_4\right)\wedge \frac{1}{2}\mathcal{F}_-\wedge \star_4\mathcal{F}_-,\label{eqn:fluxedaction}\\
=&-\mu_7 \int_{\Bbb{R}^{1,3}\times D} \text{Vol}_{\Bbb{R}^{1,3}} \wedge \mathcal{F}_-\wedge\star_4\mathcal{F}_-.
\end{align}
In the last equality we used $h^{-1}=\alpha$, i.e.~$ \text{Vol}_{\Bbb{R}^{1,3}}=C_4|_{\Bbb{R}^{1,3}}$, which holds in an ISD background.

\subsection{Inflaton potential from induced charge}

Now suppose that the D7-brane position $z_3$ is a modulus in the absence of fluxes, i.e.~suppose that $[D]\in H_4(X,\mathbb{Z})$ has a continuous family of representatives parameterized by $z_3$, which we write as $D(z_3)$.
Displacing such a D7-brane in a background of three-form flux causes ASD flux to accumulate on the D7-brane worldvolume, as we will review below.
This ASD flux carries anti-D3-brane charge, which interacts with the dissolved D3-brane charge carried by the background flux, and creates a potential for D7-brane motion.  From \eqref{eqn:fluxedaction}, this potential is
\begin{align}
V(z_3)= \mu_7 \int_{D(z_3)} h^{-1} \mathcal{F}_-\wedge\star_4\mathcal{F}_-.\label{eqn:potential and d3 charge}
\end{align}
In the special case that $h^{-1}$ is a constant,
we have
\begin{align}
V(z_3)=& 2\mu_3  h^{-1} \frac{\mu_7}{\mu_3}\int_{D(z_3)} \frac{1}{2}\mathcal{F}_-\wedge\star_4\mathcal{F}_-,\\
=& 2\mu_3 h^{-1} Q^{\overline{D3}}(z_3),\label{eqn:potential and d3 chargesimple}
\end{align}
Thus, the inflaton potential
is proportional to the induced anti-D3-brane charge.

In the simplest incarnation of Higgs-otic inflation, $D(z_3)$ is a family of effective divisors --- i.e., a D7-brane rather than an anti-D7-brane wraps $D(z_3)$ --- and the flux that accumulates on the D7-brane is ASD, corresponding to anti-D3-brane charge.  The inflaton potential in the probe approximation, and prior to including the effects of moduli stabilization, is given by \eqref{eqn:potential and d3 charge}.  At the minimum of this potential, the induced ASD flux vanishes, and the D7-brane preserves the same supersymmetry as the background $(2,1)$ flux.  A system of this sort provides a realization of F-term axion monodromy inflation \cite{Marchesano:2014mla} in string theory \cite{Ibanez:2014swa}.

In this note we will demonstrate that the relation (\ref{eqn:potential and d3 charge}) presents a strong constraint on model-building.  We will see that as a D7-brane moves one or more times around a one-cycle,
the backreaction of accumulated anti-D3-brane charge on the compactification geometry is large and rapidly changing, precluding inflation.

\subsection{An example}

A prototypical example of Higgs-otic inflation given in \cite{Ibanez:2014swa} occurs in a toroidal orientifold for which the covering orbifold is of the form $(T^4\times T^2)/\Bbb{Z}_4,$ with the orbifold action
\begin{equation}\label{eqn:Higgsotic orbifold action}
\theta:(z_1,z_2,z_3)\mapsto (-iz_1,-iz_2,-z_3).
\end{equation}
No explicit orientifold action was given in \cite{Ibanez:2014swa}.  In this section, we will take the orientifold action to be
\begin{equation}\label{eqn:Higgsotic orientifold action}
\sigma:(z_1,z_2,z_3)\mapsto (z_1,z_2,-z_3).
\end{equation}
This orientifold action is consistent with the presence of D7-branes and O7-planes whose position is described by the coordinate $z_3.$ As $\theta^2\sigma :(z_1,z_2,z_3)\mapsto -(z_1,z_2,z_3),$ another choice of orientifold action,
\begin{equation}
\sigma':(z_1,z_2,z_3)\mapsto-(z_1,z_2,z_3)\,,
\end{equation}
is
equivalent
to (\ref{eqn:Higgsotic orientifold action}).

The constant ISD fluxes allowed by the orbifold action (\ref{eqn:Higgsotic orbifold action}) are
\begin{equation}
G_+=G^{(2,1)}dz^1\wedge dz^2\wedge d\bar{z}^3+G^{(0,3)}d\bar{z}^1\wedge d\bar{z}^2\wedge d\bar{z}^3.
\end{equation}
The NS-NS three-form flux is
\begin{equation}
H=\frac{i}{2\im\tau}\Bigl(dz^1\wedge dz^2\wedge( G^{(2,1)}d\bar{z}^3-G^{(0,3)*}dz^3) +d\bar{z}^1\wedge d\bar{z}^2\wedge(G^{(0,3)}d\bar{z}^3-G^{(2,1)*}dz^3 )\Bigr).
\end{equation}
We can choose a gauge (corresponding to the normal coordinate expansion in \cite{Jockers:2004yj}) so that the NS-NS two-form field $B$ is
\begin{equation}\label{eqn:NSNS 2 form}
B=\frac{i}{2\im\tau}\Bigl(dz^1\wedge dz^2( G^{(2,1)}\bar{z}_3-G^{(0,3)*}z_3) +d\bar{z}^1\wedge d\bar{z}^2(G^{(0,3)}\bar{z}_3-G^{(2,1)*}z_3 )\Bigr).
\end{equation}
If the background \eqref{eqn:NSNS 2 form} pulled back to a D7-brane leads to ASD flux $\mathcal{F}$, then the key ingredients for Higgs-otic inflation are present.

\subsection{An issue of orientation}\label{sec:ori}

We now explain a subtlety concerning orientation and the self-duality of flux.
The most straightforward realization of the Higgs-otic scenario requires a flux background in which ASD flux is induced on a D7-brane that wraps a four-cycle $D$.
However, we will show that a $B$-field of Hodge type $(0,2)+(2,0)$, such as (\ref{eqn:NSNS 2 form}), is SD, not ASD, when $D$ is an effective divisor.

If one provisionally takes the orientation of $D$ to be
\begin{equation}\label{eqn:wrong orientation}
dz^1\wedge d\bar{z}^1\wedge dz^2\wedge d\bar{z}^2,
\end{equation}
then a $B$-field of Hodge type $(0,2)+(2,0)$ is indeed ASD, as desired for Higgs-otic inflation.  A simple check of the anti-self-duality is that $B\wedge B$ is negative relative to the orientation (\ref{eqn:wrong orientation}), as required for an ASD real two-form --- see
(\ref{eqn:antiself}).

However, we will now argue that the correct orientation for an effective divisor differs from (\ref{eqn:wrong orientation}) by a sign: as recognized in \cite{Ruehle:2017one}, the orientation (\ref{eqn:wrong orientation}) corresponds to the orientation on an anti-D7-brane, not a D7-brane, wrapping $D$.

Suppose that $X$ is a K\"ahler threefold with Hermitian metric $i\,g_{a\bar{b}}$, and let $D$ be an effective divisor written as $\{z_3=a\}$ in local coordinates.
We show in Appendix \ref{app:convention} that there are two possible choices of conventions for the Hodge star map, and correspondingly there are two choices of K\"ahler form,
which in a unitary frame read
\begin{equation}\label{eqn:kahler forms}
 J=\pm i(g_{1\bar{1}}dz^1\wedge d\bar{z}^1+g_{2\bar{2}}dz^2\wedge d\bar{z}^2)\,.
\end{equation}
Given either K\"ahler form in (\ref{eqn:kahler forms}), the volume form of $D$ is
\begin{equation}\label{eqn:right orientation}
\frac{1}{2}J\wedge J=-g_{1\bar{1}}g_{2\bar{2}}dz^1\wedge d\bar{z}^1 \wedge dz^2\wedge d\bar{z}^2.
\end{equation}
The orientation (\ref{eqn:wrong orientation}) used in \cite{Ibanez:2014swa} has opposite sign relative to (\ref{eqn:right orientation}). This implies that the volume of $D$ with the orientation (\ref{eqn:wrong orientation}) measured by the K\"ahler form (\ref{eqn:kahler forms}) is negative.
Note also that the eigenvalues of the four-dimensional Hodge star operator on $D$ change sign under a change of the sign of the volume form. As a result, the NS-NS 2-form $B$ (\ref{eqn:NSNS 2 form}), of Hodge type $(2,0)+(0,2)$, corresponds to a \emph{self-dual} 2-form given the orientation (\ref{eqn:right orientation}).

We conclude that in the particular orbifold proposed in \cite{Ibanez:2014swa}, the three-form fluxes allowed by the orbifold action \eqref{eqn:Higgsotic orbifold action} result from an NS-NS two-form $B$ \eqref{eqn:NSNS 2 form}
of Hodge type $(0,2)+(2,0)$.  Such a form is SD when pulled back to a D7-brane.\footnote{We have argued above, and in more detail in Appendix \ref{app:convention}, that the orientation of the worldvolume of a D7-brane is given by (\ref{eqn:right orientation}), which differs by a sign from the orientation (\ref{eqn:wrong orientation}) used in \cite{Ibanez:2014swa}.  Our choice of conventions is anchored by the requirement, almost ubiquitous in the literature, that $G_3$ flux of Hodge type $(0,3)$ should be ISD rather than IASD.}  We therefore find that a D7-brane displaced in the $z_3$ direction in the compactification proposed in \cite{Ibanez:2014swa}, taking \eqref{eqn:Higgsotic orientifold action} to be the orientifold action, does not accumulate ASD flux, and does not lead to axion monodromy inflation.  We have not found an alternative orientifold action that leads to a successful model based on the orbifold \eqref{eqn:Higgsotic orbifold action}.

However, we now give an example of a toroidal orientifold that could support Higgs-otic inflation. Consider the toroidal orientifold $T^6/\Bbb{Z}_6'$ studied in \cite{Berg:2004ek}, T-dualized six times in order to obtain O3-planes and O7-planes rather than O5-planes and O9-planes.
The orbifold action $\theta$ and the orientifold action $\sigma$ are
\begin{align}
&\theta:(z_1,z_2,z_3)\mapsto (e^{i\pi/3}z_1,e^{-i\pi}z_2,e^{2\pi i/3}z_3)\label{eqn:new orbifolding},\\
&\sigma:(z_1,z_2,z_3)\mapsto -(z_1,z_2,z_3) \label{eqn:new orientifolding}.
\end{align}
As $\theta^3\sigma:(z_1,z_2,z_3)\mapsto (z_1,z_2,-z_3),$ the position modulus of an inflationary D7-brane is $z_3.$   The orbifold action (\ref{eqn:new orbifolding}) allows the bulk three-form flux
\begin{equation}
G= G^{(2,1)} dz^1\wedge d\bar{z}^2\wedge dz^3,
\end{equation}
which generates an ASD $B$-field on the divisor $\{z_3=a\}$:
\begin{equation}
B=\frac{i g_s}{2} \left( G^{(2,1)}z_3 dz^1\wedge d\bar{z}^2-G^{(2,1)*}\bar{z}_3 d\bar{z}^1\wedge dz^2\right).
\end{equation}
Thus the toroidal orientifold defined by (\ref{eqn:new orbifolding}), (\ref{eqn:new orientifolding}) could support a Higgs-otic inflation scenario. However, in the presence of bulk flux of Hodge type $(0,3)$, which is required to induce a nonvanishing flux superpotential, the $(2,0)+(0,2)$ components of $\mathcal{F}$ do not vanish in general, and so the $B$ field on the divisor is a linear combination of SD and ASD components.
This leads to somewhat more complicated backreaction effects than purely ASD flux would produce, as we shall see.

\section{Backreaction of Monodromy Charge} \label{sec:backreaction}

Having recalled the essential elements of Higgs-otic inflation, most notably the contribution \eqref{eqn:potential and d3 charge} of ASD flux on the inflationary D7-brane to the inflaton potential, we can now study Higgs-otic inflation beyond the probe approximation.  We will find that the accumulation of ASD flux sources significant changes in the supergravity solution for the internal space --- changes that are omitted by assumption in the probe approximation.

In particular, we will see that the actions of Euclidean D3-branes, even those that are  well-separated from the inflationary D7-brane, depend sensitively on the inflaton vev once backreaction is included.  As a result, we will be able to draw strong conclusions about Higgs-otic inflation scenarios in which nonperturbative superpotential terms from Euclidean D3-branes\footnote{Precisely parallel results hold for superpotentials from gaugino condensation on D7-branes, but for simplicity of language we suppress the gaugino condensate case in our discussion.}
make important contributions to the potential for
the K\"ahler moduli, as in \cite{Kachru:2003aw,Balasubramanian:2005zx,Bobkov:2010rf}.
The presence of perturbative contributions to the K\"ahler moduli potential, as in the Large Volume Scenario, does not affect our conclusion: all that matters is that the nonperturbative terms play a non-negligible role in moduli stabilization.  On the flip side, our analysis does not directly constrain a hypothetical Higgs-otic inflation scenario stabilized by purely perturbative effects.

Although our computation will occur in ten-dimensional type IIB supergravity in the presence of localized and distributed sources, the results are efficiently expressed in four-dimensional $\mathcal{N}=1$ supergravity, with the superpotential
\begin{equation}
W = \int_X G\wedge \Omega + \sum_{a} \mathcal{A}_a e^{-2\pi Q_a^{~i} T_i}\,.
\end{equation}
Here $\{T_i\}$ are the complexified K\"ahler moduli, $i=1,\ldots,h^{1,1}(X)$, and the coefficients
$Q_a^{~i} \in \mathbb{Z}$ are the charges of Euclidean D3-branes under the shift symmetries of the Ramond-Ramond four-form axions.
Determining which homology classes $[D]\in H_4(X,\mathbb{Z})$ support Euclidean D3-brane superpotential terms is beyond the scope of this work, and so we do not specify the $Q_a^{~i}$ or the range of the index $a$.  It will suffice, in fact, to examine a single term, so we write
\begin{equation}
W = \int_X G\wedge \Omega +  \mathcal{A}\,e^{-2\pi T}
\end{equation} henceforth.
The Pfaffian prefactor $\mathcal{A}$ depends on the complex structure moduli, on the positions of any D3-branes \cite{Ganor:1996pe,Kachru:2003sx,Berg:2004ek,Baumann:2006th}, and, as we shall now show, \emph{on the positions of magnetized D7-branes}.

Consider a Euclidean D3-brane wrapping a holomorphic divisor $D$ in a general flux background.  No essential generality is lost in assuming that the complexified volume of $D$ is one of the K\"ahler moduli, denoted $T$.  We allow ASD flux $\mathcal{F}_D$ on the Euclidean D3-brane in accordance with the conditions for an instanton to preserve supersymmetry \cite{Marino:1999af,Bianchi:2011qh}.\footnote{Notice that on a spacetime-filling D7-brane SD flux can be supersymmetric, while on a Euclidean D3-brane only ASD flux can be supersymmetric.} The DBI action of such a magnetized Euclidean D3-brane is
\begin{align}\label{sdbi}
S_{DBI}=&\mu_3\int_D \frac{1}{2}\bigl(\mathcal{J}\wedge \mathcal{J}+\im\tau \mathcal{F}_D\wedge\star_4\mathcal{F}_D\bigr),\\
=&\mu_3\int_D \frac{1}{2}\bigl(\mathcal{J}\wedge \mathcal{J}-\im\tau \mathcal{F}_D\wedge\mathcal{F}_D\bigr).\label{sdbi2}
\end{align}
One immediate observation is that the flux-induced D(-1)-brane charge $\frac{\mu_3}{\mu_{-1}} \int_D \frac{1}{2}\mathcal{F}_D\wedge\star_4\mathcal{F}_D$ is coupled to the axio-dilaton, and so the magnetized Euclidean D3-brane should be sensitive to the D7-brane position moduli in general.

The magnitude of the Euclidean D3-brane superpotential
obeys
\begin{equation} \label{pfaffdef}
\bigl|\mathcal{A}  e^{-2\pi T}\bigr| \propto e^{-S_{DBI}}\,.
\end{equation}
One can therefore compute the Pfaffian $\mathcal{A}$ by computing $S_{DBI}$, as in \cite{Baumann:2006th}.
We will now do so to leading order in expansion around an ISD background.

\subsection{Perturbative computation of backreaction}

We begin with the full equations of motion.  Taking the ansatz \eqref{eqn:metric ansatz} and defining the quantities
\begin{align}
&\Phi_{\pm}=h^{-1}\pm\alpha,\\
&\Lambda=h^{-1}\star_6 G_3-i\alpha G_3=\Phi_+G_-+\Phi_-G_+,
\end{align}
the type IIB supergravity action \eqref{sugraaction} leads to the following equations of motion and Bianchi identities, in the conventions of \cite{Baumann:2010sx,Gandhi:2011id}:
\begin{align}
&\nabla^2\Phi_{\pm}=\frac{(\Phi_++\Phi_-)^2}{24\,\im\tau}G_{\pm,abc} \bar{G}_{\pm}^{abc}+\frac{2}{\Phi_++\Phi_-}\nabla_a\Phi_\pm \nabla^a \Phi_\pm+\kappa_{10}^2\frac{(\Phi_++\Phi_-)^2}{2}\left(\frac{1}{4}(\hat{T}^{i}_i-\hat{T}_{\mu}^{\mu})\pm \mu_3\rho^{D3} \right),\label{eqn:warping PDE}\\
&\left(d\Lambda+\frac{id\tau}{\im\tau}\wedge\re\Lambda\right)\wedge dx^0\wedge dx^1\wedge dx^2\wedge dx^3=2i\kappa_{10}^2C_4\wedge \frac{\delta S_{loc}}{\delta C_6}+2i\kappa_{10}^2\frac{\delta S_{loc}}{\delta B_2},\\
&d(G_3+\tau H_3)=dF_3=-2\kappa_{10}^2\frac{\delta S_{loc}}{\delta C_6},\\
&\nabla^2 \tau=\frac{\nabla \tau\cdot \nabla\tau}{i\im\tau}-\frac{i(\Phi_++\Phi_-)}{12}G_{+,abc} G_-^{abc}+4i\kappa_{10}^2(\im\tau)^2\frac{\delta S_{loc}}{\delta \bar{\tau}},\label{eqn:eom tau}\\
&R_{mn}=\frac{\nabla_{(m}\tau\nabla_{n)}\bar{\tau}}{2(\im\tau)^2}+\frac{2}{(\Phi_++\Phi_-)^2}\nabla_{(m}\Phi_+\nabla_{n)}\Phi_--g_{mn}\frac{\mathcal{R}_4}{2(\Phi_++\Phi_-)}\nonumber\\
&\qquad\qquad-\frac{\Phi_++\Phi_-}{8\im\tau}\left(G_{+(m}^{~~~~pq}\bar{G}_{-n)pq}+G_{-(m}^{~~~~pq}\bar{G}_{+n)pq} \right)+\kappa_{10}^2\left(\hat{T}_{mn}-\frac{1}{4}g_{mn}\hat{T}^{i}_i\right),\label{eqn:warping PDEend}
\end{align}
where $\hat{T}$ is the energy momentum tensor of localized objects such as D-branes and O-planes.

\subsubsection{Approximation scheme and simplifying assumptions}

We would like to solve the system \eqref{eqn:warping PDE}-\eqref{eqn:warping PDEend} to leading order in the effects of the two-form flux $\mathcal{F}$ that accumulates on the inflationary D7-brane.

To this end, we consider a compactification of type IIB superstring theory on a toroidal orientifold\footnote{The toroidal orientifold restriction makes it possible to compute the explicit Green's function, see \eqref{eqn:6d green}. We expect, but will not show here, that our qualitative results hold more generally.} with local coordinates $(z_1,z_2,z_3)$,
containing
O7-planes, magnetized D7-branes, ISD flux, O3-planes, and possibly also D3-branes.
We will first find a background solution containing ISD flux, O3-planes, and --- optionally --- D3-branes, with $\Phi_-=0$.  Then we will perturb the equations of motion by including the O7-planes and magnetized D7-branes as localized source terms.

Without loss of generality, we assume that the orientifold involution is $\sigma :z_3\mapsto -z_3,$ so that the O7-planes and D7-branes are extended over the $z_1$ and $z_2$ directions.
We assume that each D7-brane $\alpha$ wraps a holomorphic divisor $D_\alpha=\{z_3=z_{3,\alpha}\},$ whose unwarped volume is $\int_D \text{Vol}_D= {\mathrm{Re}}\,T_D.$
The D7-brane charge density is then
\begin{equation}\label{d7rho}
\rho^{D7}(z_3)=\sum_\alpha \rho^{D7}_\alpha \delta_{(2)}(z_3-z_{3,\alpha}),
\end{equation}
where $\rho^{D7}=1$ for D7-branes and $\rho^{D7}=-4$ for O7-planes. Because we have assumed that the background ISD flux includes nonzero components of Hodge types $(0,3)$ and $(2,1)$, the two-form flux $\mathcal{F}$ on a D7-brane may include both ASD and SD components --- see \eqref{eqn:general 2 flux}.  We do not consider any flux on the O7-planes.

The D3-brane charge density of D3-branes and O3-planes takes the form
\begin{equation}
\rho^{D3}(z)=\sum_i \rho^{D3}_i \delta_{(6)}(z-z_i),\label{eqn:local D3 charge}
\end{equation}
where $z_i$ is the position of the D3-brane or O3-plane, $\rho^{D3}_i=1$ for D3-branes, and $\rho^{D3}_i=-1/4$ for O3-planes.

A primary focus of this note is the DBI action \eqref{sdbi} of a Euclidean D3-brane at a fixed location. The NS-NS two-form $B$ pulled back to the Euclidean D3-brane describes how NS-NS three-form flux $H$ accumulates under a displacement of the Euclidean brane along the normal direction.  Thus for a Euclidean D3-brane at a fixed location, corrections to $H$ do not significantly affect the DBI action \eqref{sdbi}. This allows us to
consider only the fields $\Phi_\pm,$ $\tau,$ and $g_{mn}$ in the perturbed equations of motion.

To achieve considerable gains in simplicity, we will only focus on \emph{localized} sources, such as those in \eqref{d7rho} and \eqref{eqn:local D3 charge}, in the perturbed equations of motion. We will find that localized stress-energy and charge associated to ASD flux on the inflationary D7-brane strongly affects the solution at other locations in the compactification, including on the divisors wrapped by Euclidean D3-branes.  While it is logically possible that including the backreaction of distributed sources, such as bulk three-form flux, could produce a counterbalancing effect on the Euclidean D3-brane action and leave the inflationary model unmodified in the final account, we find such a conspiracy to be most implausible.

Away from the minimum of the inflaton potential, the energy stored in the D7-brane configuration presents an obstacle to solving the ten-dimensional equations of motion with purely classical sources.
We refer to such an obstacle as an NS-NS tadpole.  In our ten-dimensional analysis we assume that there exist sources that cancel all NS-NS tadpoles, i.e.~we assume that perturbative and nonperturbative corrections to the ten-dimensional equations of motion allow for consistent cosmological solutions.
One leading candidate for an effect that cancels NS-NS tadpoles is gaugino condensation, as in \cite{Baumann:2010sx,Moritz:2017xto}, but establishing NS-NS tadpole cancellation from specific quantum effects is beyond the scope of this work.

Practically, for a bosonic supergravity field $A,$ we expand $A=A^{(0)}+A^{(1)}+\cdots,$ where $A^{(0)}$ is the background field, and $A^{(1)}$ is the perturbed field at leading order. Given this expansion, we rewrite the perturbed equations of motion schematically as
\begin{equation}
\nabla^2 A^{(1)}=\rho^{bulk}_{A}+\rho_A^{D7}+\rho_A',
\end{equation}
where $\rho^{bulk}_{A}$ is a bulk source term that involves bulk fields, $\rho_A^{D7}$ is a source term that is localized on D7-branes, and $\rho_A'$ is a source term added by hand to ensure tadpole cancellation. We write $\rho_A^{NS}:=\rho^{bulk}_{A}+\rho_A'$ and refer to $\rho_A^{NS}$ as the NS-NS tadpole cancelling source.

As an example, we expand the equation of motion of $\tau.$ The kinetic term $\nabla^2 \tau$ is expanded as
\begin{equation}
\left( \nabla^2\tau\right)^{(1)}=\nabla^{2(0)}\tau^{(1)}+\nabla^{2(1)}\tau^{(0)},
\end{equation}
where we often write $\nabla^{2(0)}$ as $\nabla^2$ when there is no ambiguity. Similarly, we expand the terms on the right hand side of \eqref{eqn:eom tau} and treat the D7-brane density as a first-order term,
\begin{equation}
\nabla^{2(0)}\tau^{(1)}+\nabla^{2(1)}\tau^{(0)}=\left( \frac{\nabla \tau\cdot\nabla\tau}{i\im\tau}\right)^{(1)}-\left(\frac{i(\Phi_++\Phi_-)\langle G_+,G_-\rangle}{2} \right)^{(1)} + 4i\kappa_{10}^2(\im\tau)^2\frac{\delta S_{loc}}{\delta \bar{\tau}}.
\end{equation}
Note that for $\tau$ we do not have to add a term by hand to ensure tadpole cancellation at leading order.
The localized source $\rho_{\tau}^{D7}$ is
\begin{equation}
\rho_\tau^{D7}=4i\kappa_{10}^2(\im\tau)^2\frac{\delta S_{loc}}{\delta \bar{\tau}}.
\end{equation}
Then we define $\rho_\tau^{bulk}$ by
\begin{equation}
\rho_\tau^{bulk}=-\nabla^{2(1)}\tau^{(0)}+\left( \frac{\nabla \tau\cdot\nabla\tau}{i\im\tau}\right)^{(1)}-\left(\frac{i(\Phi_++\Phi_-)\langle G_+,G_-\rangle}{2} \right)^{(1)},
\end{equation}
and $\rho_\tau^{bulk}$ is identical to $\rho_\tau^{NS}$ due to the absence of a $\rho_\tau'$ term. Finally, we write down the perturbed equation of motion for $\tau$ as
\begin{equation}
\nabla^2\tau^{(1)}=\rho_\tau^{D7}+\rho_\tau^{NS}.
\end{equation}
For further details of this perturbation scheme, see \cite{Gandhi:2011id}.

Strictly speaking, the perturbed dilaton could be negative in a small region around an O7-plane,
and the perturbed metric could be negative around a magnetized D7-brane.  To suppress effects of these singular regions on our solution, we will require $\Phi_{+,c} \ll g_s \ll 1$.

\subsubsection{Perturbed equations of motion}\label{section:perturbed equations of motion}

Consider first a compactification containing only ISD flux, D3-branes, and O3-planes, so that $\Phi_-^{(0)}=\Lambda^{(0)}=0.$
With the localized source terms \eqref{eqn:local D3 charge}, the equations of motion and Bianchi identities \eqref{eqn:warping PDE} are
\begin{align}
&d\Bigl(G_3^{(0)}+\tau^{(0)} H_3^{(0)}\Bigl)=0,\\
&\nabla^2 \tau^{(0)}=\frac{\nabla\tau^{(0)}\cdot\nabla\tau^{(0)}}{i\im\tau^{(0)}},\\
&R_{mn}^{(0)}=\frac{\nabla_{(m}\tau^{(0)}\nabla_{n)}\bar{\tau}^{(0)}}{2(\im\tau^{(0)})^2}+\frac{2}{\Phi_+^{(0)}+\Phi_-^{(0)}}\nabla_{(m}\Phi_+^{(0)}\nabla_{n)}\Phi_-^{(0)},\\
&\nabla^2(\Phi_+^{(0)})^{-1}=-\sum_i \mu_3\kappa_{10}^2\rho_i^{D3} \delta_{(6)}(z-z_i)-\frac{|G_+^{(0)}|^2}{4\im\tau^{(0)}}.\label{eqn:D3 brane charge eom}
\end{align}
The solutions for the ISD background are then
\begin{align}
&(\Phi_+^{(0)})^{-1}=\Phi_{+,c}^{-1}-\sum_i \mu_3\kappa_{10}^2 \rho_i^{D3} G_{(6)}(z;z_i),\\
&\tau^{(0)}=i/g_s,\\
&g_{z_1\bar{z}_1}^{(0)}=g_{z_2\bar{z}_2}^{(0)}=g_{z_3\bar{z}_3}^{(0)}=1/2,
\end{align}
where $\Phi_{+,c}^{-1}$ and $g_s$ are constants.  Denoting the unwarped volume of the compactification by $\mathcal{V}$, the warped volume $\mathcal{V}_{w}$ is then
\begin{equation}
\mathcal{V}_{w} = \left(\frac{\Phi_{+,c}^{-1}}{2}\right)^{3/2}\mathcal{V}.
\end{equation}
The D3-brane charge dissolved in ISD flux is
\begin{equation}
Q^{D3}_{flux}=\frac{|G_+^{(0)}|^2\mathcal{V}}{4\mu_3\kappa_{10}^2\im\tau^{(0)}}.\label{eqn:bulk d3 charge}
\end{equation}

Now we incorporate localized magnetized D7-branes, as well as O7-planes, as perturbations of the above background.  The perturbed equations of motion are
\begin{align} \label{perteom1}
&\nabla^2\Phi_-^{(1)}=-\frac{1}{4}\sum_\alpha \mu_7\kappa_{10}^2\rho_\alpha^{D7}\Phi_+^{(0)2}\tr(g^{(0)-1}\mathcal{F}_{-,\alpha})^2\delta_{(2)}(z_3-z_{3,\alpha})+\rho_-^{NS},\\ \label{perteom2}
&\nabla^2(\Phi_+^{(1)})^{-1}=\frac{1}{4}\sum_\alpha \mu_7\kappa_{10}^2\rho_\alpha^{D7}\tr(g^{(0)-1}\mathcal{F}_{+,\alpha})^2\delta_{(2)}(z_3-z_{3,\alpha})+\rho_+^{NS},\\
&\nabla^2\im\tau^{(1)}=-2\mu_7\kappa_{10}^2\sum_\alpha \rho_\alpha^{D7}\delta_{(2)}(z_3-z_{3,\alpha})+\rho_{\im\tau}^{NS}, \label{perteom3}
\end{align}
\begin{align}
\Delta_K g_{mn}^{(1)}=&\rho_{g,mn}^{NS}(z_3)-2\mu_7\kappa_{10}^2(\im\tau^{(0)})^{-1}\sum_\alpha \rho^{D7}_\alpha \delta_{(2)}(z_3-z_{3,\alpha})\delta_{z_3(m}\delta_{n)\bar{z}_3}\nonumber\\
&+\mu_7\kappa_{10}^2\sum_\alpha\Phi^{(0)}_+ \rho_\alpha^{D7}\delta_{(2)}(z_3-z_{3,\alpha})\Bigl(\mathcal{F}_{ma,\alpha}\mathcal{F}_{nb,\alpha}g^{(0)ab}-\frac{1}{2}g_{mn}^{||(0)}|\mathcal{F}_\alpha|^2\Bigr), \label{perteom4}
\end{align}
where
$\rho_-,$ $\rho_+,$ $\rho_{\im\tau}$, and $\rho_{g}$ are NS-NS tadpole cancelling sources, $g_{mn}^{||(0)}$ is the background metric with legs parallel to the D7-brane divisor, and
\begin{equation}
\Delta_K g_{mn}^{(1)}:=\nabla^2g_{mn}^{(1)}+\nabla_m\nabla_ng^{(0)ab}g_{ab}^{(1)}.
\end{equation}
Equation \eqref{perteom4} can be separated into two equations,
\begin{equation}
\nabla^2g^{(0)ab}g_{ab}^{(1)}=-4\mu_7\kappa_{10}^2(\im\tau^{(0)})^{-1}\sum_\alpha \rho_\alpha^{D7}\delta_{(2)}(z_3-z_{3,\alpha})+g^{(0)ab}\rho_{g,ab},\label{perteom5}
\end{equation}
\begin{equation}
\nabla^2g_{mn}^{||(1)}=\mu_7\kappa_{10}^2\sum_\alpha\Phi^{(0)}_+(z_{3,\alpha}) \rho_\alpha^{D7}\delta_{(2)}(z_3-z_{3,\alpha}) \Bigl(\mathcal{F}_{ma,\alpha}\mathcal{F}_{nb,\alpha}g^{(0)ab}-\frac{1}{2}g_{mn}^{||(0)}|\mathcal{F}_\alpha|^2\Bigr)+\rho_{g,mn}^{NS,||}(z_3).\label{perteom6}
\end{equation}
\subsubsection{Solution incorporating backreaction}

The solutions
for the equations \eqref{perteom1}-\eqref{perteom3} and \eqref{perteom5}-\eqref{perteom6}
are readily obtained in terms of the scalar Green's functions $G_{(6)}(z;z')$ and $G_{(2)}(z_3;z_{3}')$ derived in Appendix \ref{app:green}:
\begin{align}\label{soln1}
\Phi_-^{(1)}(z)=&-\frac{1}{4}\sum_\alpha \mu_7\kappa_{10}^2\rho_\alpha^{D7}\int_D d^4z' G_{(6)}(z;z') \Phi_+^{(0)2}(z',z'_{3,\alpha})\tr(g^{(0)-1}\mathcal{F}_{-,\alpha})^2\nonumber\\&+\int_X G_{(6)}(z;z')\rho^{NS}_-(z'),
\end{align}
\begin{equation}\label{soln2}
(\Phi_+^{(1)})^{-1}(z)=\frac{1}{4}\sum_\alpha \mu_7\kappa_{10}^2\rho_\alpha^{D7}G_{(2)}(z_3;z'_{3,\alpha})\tr(g^{(0)-1}\mathcal{F}_{+,\alpha})^2+\int_X G_{(6)}(z;z')\rho^{NS}_+(z'),
\end{equation}
\begin{equation}\label{soln3}
\im\tau^{(1)}(z_3)=-2\mu_7\kappa_{10}^2\sum_\alpha \rho^{D7}_\alpha G_{(2)}(z_3;z_{3,\alpha})+\int_X G_{(6)}(z;z')\rho^{NS}_{\im\tau}(z'),
\end{equation}
\begin{equation}\label{soln4}
g^{(0)ab}g_{ab}^{(1)}=-4\mu_7\kappa_{10}^2(\im\tau^{(0)})^{-1}\sum_\alpha \rho_{\alpha}^{D7}G_{(2)}(z_3;z_{3,\alpha})+\int_{D^\perp} G_{(2)}(z_3;z'_3) g^{(0)ab}\rho^{NS}_{R,ab},
\end{equation}
\begin{align}\label{soln5}
g_{mn}^{||(1)}(z_3)=&\mu_7\kappa_{10}^2\sum_\alpha \Phi^{(0)}_+(z_{3,\alpha})\rho_\alpha^{D7}G_{(2)}(z_3;z_{3,\alpha}) \Bigl(\mathcal{F}_{ma,\alpha}\mathcal{F}_{nb,\alpha}g^{(0)ab}-\frac{1}{2}g_{mn}^{||(0)}|\mathcal{F}_\alpha|^2\Bigr)\nonumber\\&+\int_{D^\perp}G_{(2)}(z_3;z'_3)\rho^{NS,||}_{g,mn}(z'_3),
\end{align} where $D^\perp$ denotes the two-cycle dual to $D.$

\subsection{Effects on Euclidean D3-branes}

Now we examine the DBI action \eqref{sdbi2} for a Euclidean D3-brane wrapping a divisor $D$ that is parallel\footnote{Our methods can also be applied when $D$ is not parallel to the $D_{\alpha}$, though we will not present the non-parallel case in this note.} to the D7-brane divisors $D_{\alpha}$.
In local coordinates, \eqref{sdbi2} can be written
\begin{equation}
S_{DBI}=\mu_3\int_D h\,g_{z_1\bar{z}_1}g_{z_2\bar{z}_2}dz^1\wedge dz^2\wedge d\bar{z}^1\wedge d\bar{z}^2-\frac{\im\tau}{2} \mathcal{F}_D\wedge \mathcal{F}_D,\label{eqn:DBi action in local coordinate chart}
\end{equation}
which to first order in the perturbations is
\begin{align}
S_{DBI}^{(1)}=&2\mu_3\int_D d^4z \left(\left(\Phi_+^{(1)}\right)^{-1}-\left(\Phi_+^{(0)}\right)^{-2} \Phi_-^{(1)}\right)+\left(\Phi_+^{(0)}\right)^{-1}\left(g_{z_1\bar{z}_1}^{(1)}g_{z_1\bar{z}_1}^{(0)-1}+g_{z_2\bar{z}_2}^{(0)-1}g_{z_2\bar{z}_2}^{(1)}\right)\nonumber\\
&-\mu_3\int_D\frac{\im\tau^{(1)}}{2}\mathcal{F}_D\wedge\mathcal{F}_D.\label{eqn:perturbed DBI}
\end{align}
Evaluated in the perturbed solution given by \eqref{soln1}-\eqref{soln5},
the DBI action \eqref{eqn:perturbed DBI} reads
\begin{align}
S_{DBI}^{(1)}=&-\mu_3\sum_\alpha \rho_\alpha^{D7}\int_{D}\left( \frac{1}{2} \mathcal{F}_{-,\alpha}\wedge \star_4\mathcal{F}_{-,\alpha}+\frac{1}{2} \mathcal{F}_{+,\alpha}\wedge \star_4\mathcal{F}_{+,\alpha}\right)G_{(2)}(z_3;z_{3,\alpha})\nonumber\\
&-\mu_3\sum_\alpha\rho_\alpha^{D7}\int_{D}\frac{1}{2} \mathcal{F}_D\wedge\star_4\mathcal{F}_D ~G_{(2)}(z_3;z_{3,\alpha}),\label{eqn:perturbed dbi result1}
\end{align}
where $G_{(2)}(z_3;z_{3,\alpha})$ is the two-dimensional Green's function (\ref{eqn:2d Green's function on toroidal orientifolds}).
If we express the induced D3-brane charge and $\overline{D3}$ brane charge as
\begin{equation}
Q^{D3}_\alpha=\frac{\mu_7}{\mu_3}\int_D\frac{1}{2} \mathcal{F}_{+,\alpha}\wedge\star_4\mathcal{F}_{+,\alpha},\label{eqn:d3 monodromy charge}
\end{equation}
\begin{equation}
Q^{\overline{D3}}_\alpha=\frac{\mu_7}{\mu_3}\int_D \frac{1}{2} \mathcal{F}_{-,\alpha}\wedge\star_4\mathcal{F}_{-,\alpha},\label{eqn:anti d3 monodromy charge}
\end{equation}
and define
\begin{equation}
Q^{\overline{D3}}_D=\frac{\mu_7}{\mu_3}\int_D \frac{1}{2} \mathcal{F}_{-,D}\wedge\star_4\mathcal{F}_{-,D},\label{eqn:anti d3 monodromy charge on D}
\end{equation}
then (\ref{eqn:perturbed dbi result1}) takes the form
\begin{equation}
S_{DBI}^{(1)}=-2\pi\sum_\alpha\rho_\alpha \left( Q_\alpha^{D3}+Q_\alpha^{\overline{D3}}+Q_D^{\overline{D3}}\right) G_{(2)}(z_3;z_{3,\alpha}).\label{eqn:perturbed dbi result2}
\end{equation}

We can now read off the effect of magnetized D7-branes on the nonperturbative superpotential.  Writing \eqref{pfaffdef} as
\begin{equation} \label{pfaffpert}
\bigl|\mathcal{A}  e^{-2\pi T}\bigr| = \mathcal{A}_0\exp\bigl(-S_{DBI}^{(0)}-S_{DBI}^{(1)}\bigr)\,,
\end{equation} and noting that $S_{DBI}^{(0)}=2\pi T-2\pi\sum_i \rho_i^{D3}G_{(2)}(z_3;z_{3,i})$, we decompose the Pfaffian factor $\mathcal{A}$ into $\mathcal{A}_0,$ $\mathcal{A}_{D3},$ and $\mathcal{A}_{\mathcal{F}}$:
\begin{equation}\label{eqn:pfaffian decomposition}
\mathcal{A}=\mathcal{A}_0\mathcal{A}_{D3}\mathcal{A}_{\mathcal{F}},
\end{equation}
where $\mathcal{A}_0$ encodes the dependence on the complex structure moduli of the internal space, $\mathcal{A}_{D3} = \exp\bigl(2\pi\sum_i \rho_i^{D3}G_{(2)}(z_3;z_{3,i}) \bigr)$ encodes the dependence on the positions $z_{3,i}$ of D3-branes, and $\mathcal{A}_{\mathcal{F}}$ encodes the dependence on the positions $z_{3,\alpha}$ of magnetized D7-branes.
From \eqref{eqn:perturbed dbi result2}, the Pfaffian factor $\mathcal{A}_\mathcal{F}$ takes the form
\begin{align}
\boxed{\vphantom{\Biggl(\Biggr)}\mathcal{A}_\mathcal{F}=\exp\left(2\pi\sum_\alpha \rho_{\alpha} \bigl(Q^{D3}_\alpha+Q^{\overline{D3}}_\alpha+Q^{\overline{D3}}_D\bigr)G_{(2)}(z_3;z_{3,\alpha})\right).}\label{eqn:the one loop pfaffian}
\end{align}
Equation (\ref{eqn:the one loop pfaffian}) is one of our main results.

The final expression (\ref{eqn:the one loop pfaffian}) is rather simple, especially in view of the intricate system of perturbed equations of motion presented in \S\ref{section:perturbed equations of motion}.  The emergent simplicity can be understood as follows.
Magnetized D7-branes can be viewed as bound states of D7-branes with D3-branes dissolved as
the flux (\ref{eqn:general 2 flux}), and one should expect the Pfaffian to depend on the position moduli of this dissolved D3-brane charge (\ref{eqn:d3 monodromy charge}), (\ref{eqn:anti d3 monodromy charge}), just as the factor $\mathcal{A}_{D3}$ depends on the positions of mobile D3-branes that are not bound to a D7-brane.  Our explicit computation shows that this expectation is precisely fulfilled.

While the terms proportional to $Q_\alpha^{D3}$ and $Q_\alpha^{\overline{D3}}$ represent the backreaction of induced D3-brane charge on the warped volume of a Euclidean D3-brane, the term involving $Q_D^{\overline{D3}}$ has a qualitatively different origin.  It encodes the change in the action of a \emph{magnetized}  Euclidean D3-brane, with magnetization $\mathcal{F}_D$, that results from the dilaton profile due to the mobile D7-branes.  With a slight abuse of language we may call $Q^{\overline{D3}}_D$ the induced D(-1)-brane charge.

Using the explicit form \eqref{eqn:2d Green's function on toroidal orientifolds} for $G_{(2)}(z_3;z_{3,\alpha})$, the Pfaffian (\ref{eqn:the one loop pfaffian}) from a single magnetized D7-brane $\alpha$ is
\begin{align}
\mathcal{A}_\mathcal{F}=&\prod_{i=1}^{N}\Biggl[\left|\vartheta_1\left(\left.\frac{z_3-\theta^i z_{3,\alpha}}{L}\right|U \right)\,\eta^{-1}(U) \right|\exp\left( \frac{-\pi \im(z_3-\theta^i z_{3,\alpha})^2}{L^2\im U}\right) \Biggr]^{Q^{D3}_\alpha+Q^{\overline{D3}}_\alpha+Q_{D}^{\overline{D3}}},\label{eqn:the one loop pfaffian explicit}
\end{align}
where $L$ is the lattice size of the torus, $U$ is the complex structure modulus of the torus, and $\theta$ is the orientifold and orbifold action.

\section{Implications}\label{sec:implications}

We have shown in \S\ref{sec:backreaction} that the nonperturbative superpotential depends on the positions of magnetized D7-branes, as in (\ref{eqn:the one loop pfaffian}) and (\ref{eqn:the one loop pfaffian explicit}), because the D3-brane charge induced on the D7-branes backreacts on the internal space.  Thus, in D7-brane monodromy models, backreaction of monodromy charge leads to inflaton-dependence of the nonperturbative terms in the moduli potential.

\subsection{Inflaton-dependence of the Pfaffian}

To understand how these couplings affect inflation, we can relate the induced charges $Q^{D3}$, $Q^{\overline{D3}}$, and $Q_{D}^{\overline{D3}}$ in \eqref{eqn:the one loop pfaffian explicit} to the position $z_{3,\alpha}$ of the inflationary D7-brane, and in turn to the canonically-normalized inflaton field $\varphi$.
From \eqref{eqn:the one loop pfaffian explicit} it is clear that unless $Q_{\rm{tot}}:={Q^{D3}_\alpha+Q^{\overline{D3}}_\alpha+Q_{D}^{\overline{D3}}}$ is very small compared to unity, the dependence of $\vartheta_1$ on $z_{3,\alpha}$ causes $\mathcal{A}_{\mathcal{F}}$ to oscillate strongly over a cycle
$z_{3,\alpha} \to z_{3,\alpha}+L$.  By definition, axion monodromy involves traversing $N>1$ periods of the axion, so the oscillations could in principle be repeated $N$ times. In practice, the change in the moduli potential after a fraction of a cycle is large enough to destabilize the configuration, for example toward decompactification.
Barring a mechanism that weakens the inflaton-dependence of the superpotential compared to what we have found, prolonged inflation --- whether small-field or large-field --- does not occur.

One could ask whether for fine-tuned values of the complex structure modulus $U$ the dependence \eqref{eqn:the one loop pfaffian explicit} might be mild enough to allow inflation.
A numerical investigation has produced no evidence for this possibility, whereas fine-tuning of $U$ can partially alleviate the eta problem \cite{Berg:2004ek,Haack:2008yb} in the related D3-D7 model \cite{Dasgupta:2002ew}.  The distinction is that in a small-field model, a problematic Pfaffian coupling matters only very near a single point in field space, such as a hilltop or inflection point, and correspondingly can sometimes be fine-tuned to vanish by adjusting a single number, such as $U$.  But for D7-brane monodromy to be possible despite the coupling \eqref{eqn:the one loop pfaffian explicit}, it would be necessary to fine-tune away the problematic terms along the entire trajectory, i.e.~over one or more complete cycles.  This is a concrete incarnation of the notorious problem of functional fine-tuning in large-field inflation.

A further perspective on our findings comes from \cite{Ruehle:2017one}, in which Ruehle and Wieck studied Pfaffian couplings in an effective supergravity theory.
They considered a K\"ahler potential and superpotential of the form
\begin{align}
&K=-3\log(T+\bar{T})+\frac{1}{2}(\Phi+\bar{\Phi})^2,\\
&W=W_0+\mu\Phi^2+A_0 \vartheta_3(i\Phi,q)^\delta e^{-\alpha T}, \label{rw2}
\end{align}
where $\Phi$ corresponds to a D7-brane position modulus, $T$ is a K\"ahler modulus, and $W_0$, $\mu$, $\alpha$, $q$, $\delta$, and $A_0$ are constants.  It was shown in \cite{Ruehle:2017one} that for $\delta \gtrsim 1/2,$ the modulation of the potential via the inflaton-dependence of the Pfaffian is strong enough to adversely affect inflation.\footnote{The results of \cite{Ruehle:2017one} accord with the general finding, in the context of D3-brane inflation models, that the displacement of even a single unit of D3-brane charge typically causes a sizable correction to the Pfaffian of the nonperturbative superpotential \cite{Kachru:2003sx,Berg:2004ek,McAllister:2005mq,Baumann:2006th}, and so precludes inflation.}
Comparing \eqref{rw2} and (\ref{eqn:the one loop pfaffian explicit}), we have $\delta = Q_{\rm{tot}}$.

To apply the results of \cite{Ruehle:2017one}, we can estimate $Q_{\rm{tot}}$.  For the benchmark values for the potential given in \cite{Ibanez:2014swa}, $V(\phi)\alpha'^2 \sim \mathcal{O}(1)$, the induced $\overline{D3}$ charge is of order
\begin{equation}\label{qd3est}
Q^{\overline{D3}}\simeq\mathcal{O}(500 h^{-1})\,.
\end{equation}
Since $h \lesssim 1$, we conclude that Pfaffian couplings due to the backreaction of induced D3-brane charge spoil Higgs-otic inflation for the benchmark parameters of \cite{Ibanez:2014swa}.

To understand how the importance of backreaction depends on compactification parameters away from these benchmark values, we examine a simplified model.
We consider the two-form flux (\ref{eqn:NSNS 2 form}) on the inflationary D7-brane divisor $D$, and we only include bulk fluxes of Hodge type $(2,1)$.\footnote{As explained in \S\ref{sec:ori}, these restrictions are problematic in complete models, but they are innocuous for the present purpose of obtaining parametric scalings.}
The self-dual two-form flux (\ref{eqn:NSNS 2 form}) induces D3-brane charge on $D$:
\begin{align}
Q^{D3}=&\frac{\mu_7 \mathrm{Re}(T)}{\mu_3} \frac{|B|^2}{2},\\
=&\frac{g_s^2 \mathrm{Re}(T)}{4\mu_3/\mu_7}|G^{(2,1)}\bar{z}|^2.\label{eqn:induced charge in higgsotic}
\end{align}
Identifying the inflaton with $\im(z),$ the induced charge (\ref{eqn:induced charge in higgsotic}) simplifies to
\begin{align}
Q^{D3}=&\frac{\mu_7 \mathrm{Re}(T) \Phi_{+,c}^{-3/2}}{2\sqrt{2}g_s M_p^2}Q_{flux}^{D3} \im(z)^2,\\
=&\frac{1}{2} g_s Q_{flux}^{D3}N_w^2,\label{eqn:winding monodromy charge}
\end{align}
where $N_w=\im(z)/L.$  In \eqref{eqn:winding monodromy charge} we used the relation $M_p^2=(\Phi_{+,c}^{-3/2} \mathcal{V})/(2\sqrt{2}g_s^2\kappa_{10}^2).$ In terms of the canonically normalized field $\varphi$, for small field excursions $\varphi\lesssim \mathcal{O}(M_{p})$ the induced charge (\ref{eqn:induced charge in higgsotic}) is given by
\begin{equation}
Q^{D3}=\frac{\Phi_{+,c}^{-1/2}}{4\sqrt{2}g_s^2}Q_{flux}^{D3}\left|\frac{\varphi}{M_p}\right|^2,
\end{equation}
whereas for large field excursions, $\varphi\gtrsim\mathcal{O}(M_p),$
\begin{equation}
Q^{D3}=\sqrt{\frac{\Phi_{+,c}^{-1/2}}{8\pi\sqrt{2}} \mu_3 \Phi_{+,c}^{-1}\mathrm{Re}(T) Q^{D3}_{flux}}\left|\frac{\varphi}{M_p}\right|.
\end{equation}
Note that $\mu_3 \Phi_{+,c}^{-1}\mathrm{Re}(T)$ is the DBI action of a Euclidean D3-brane wrapping $D$.

To display the leading dependence of the Pfaffian \eqref{eqn:the one loop pfaffian} on $\varphi,$ we make further simplifications: we set $U \to 1$, we omit the orientifold images of the magnetized D7- branes, and we expand $\vartheta_1$ for small displacements $z_3/L \ll 1$.
This yields
\begin{equation}
\mathcal{A}_{\mathcal{F}}(\varphi)\simeq \left[ c \frac{\varphi-\varphi_0}{M_p}\exp\left(- \pi c^2\frac{(\varphi-\varphi_0)^2}{M_p^2}\right)\right]^{d\frac{\varphi^2}{M_p^2}},\label{eqn:the one loop pfaffian higgsotic}
\end{equation}
where $c=\Phi_{+,c}^{-1/4}/(2^{3/4} g_s^{3/2})$, $d=\frac{1}{2}g_s c^2 Q_{flux}^{D3}$, and $\varphi_0$ is the location of the Euclidean D3-brane, expressed in terms of the canonically-normalized D7-brane position coordinate $\varphi$.
For $g_s \ll 1$ and $Q_{flux}^{D3} \gg 1$ we have $c \gg 1$ and $d \gg 1$, and even for the marginally controllable parameter choice $g_s = 1/2$, $Q_{flux}^{D3}=1$ we have $c>1$ and $d\simeq0.7$. Because the equations \eqref{eqn:D3 brane charge eom} and \eqref{eqn:bulk d3 charge} imply that $Q_{flux}^{D3}$ is integrally quantized, $d$ cannot be made arbitrarily small for $g_s<1$.
Evidently the Pfaffian \eqref{eqn:the one loop pfaffian} cannot be approximated by a constant independent of $\varphi$.

\subsection{Comment on fluxbrane inflation}

Even though the primary focus of this note has been on the backreaction of monodromy charge in the Higgs-otic model, the dependence of the Pfaffian (\ref{eqn:the one loop pfaffian}) on the induced charge (\ref{eqn:d3 monodromy charge}), (\ref{eqn:anti d3 monodromy charge}) has broader applicability. We now discuss the implications of (\ref{eqn:the one loop pfaffian}) for fluxbrane inflation.

Fluxbrane inflation \cite{Hebecker:2011hk, Hebecker:2012aw, Arends:2014qca} is a hybrid inflation scenario in string theory in which the inflaton field is the separation
of a pair of spacetime-filling D7-branes.
Suppose that $X$ is an orientifold of a Calabi-Yau threefold, with $[\Sigma] \in H_4(X,\mathbb{Z})$ a homology class that admits a continuous family of holomorphic representatives.
Two D7-branes $\mathcal{D}_a$ and $\mathcal{D}_b$ can then be wrapped on distinct representatives $\Sigma_a, \Sigma_b \in [\Sigma]$. The proposal of \cite{Hebecker:2011hk} was to introduce a non-supersymmetric relative gauge flux $\mathcal{F}$ on $\mathcal{D}_a$ and $\mathcal{D}_b$, so that the D7-branes feel an attractive force and are driven to meet and fuse.

In order for inflation to be possible in this scenario, the flux $\mathcal{F}$ must fulfill certain conditions.
First, $\mathcal{F}$ should be chosen to lie in the part of $H^2(\Sigma)$ that descends from $H^2(X)$: this ensures the absence of a superpotential term of the form $\int_{\mathcal{C}_5} \Omega \wedge \mathcal{F}$, with $\mathcal{C}_5$ a five-chain ending on $\Sigma$.  If such a term were present it could produce a problematically large F-term potential for the D7-brane position, cf.~\cite{Jockers:2005zy}.
Next, some choices of $\mathcal{F}$ will induce D3-brane charge on the D7-branes, and it is well-known that such D3-brane charge can lead to significant couplings in the nonperturbative superpotential \cite{Kachru:2003sx,Berg:2004ek,Baumann:2006th,Marchesano:2009rz}.
In order to avoid unwanted forces from induced D3-brane charge, the authors of \cite{Hebecker:2012aw} imposed the requirement
\begin{equation}\label{zerod3tot}
\int_\Sigma \mathcal{F}\wedge \mathcal{F}=0\,.
\end{equation}
Because $\int_\Sigma\mathcal{F}\wedge\mathcal{F}=\int_\Sigma\mathcal{F}_{+}\wedge\star_4\mathcal{F}_{+}-\int_\Sigma\mathcal{F}_{-}\wedge\star_4\mathcal{F}_{-}$,
the condition \eqref{zerod3tot} enforces that the \emph{net} induced D3-brane charge
vanishes, but allows D3-brane and anti-D3-brane charge density to be present in equal amounts.  Thus, imposing \eqref{zerod3tot} does not suffice to ensure that the backreaction of D3-brane charge vanishes: the SD and ASD components separately provide source terms.

Let us therefore examine the backreaction of induced charge on the Pfaffian in fluxbrane inflation.
The induced D3 brane tension
\begin{equation}
\frac{\mu_7}{\mu_3} \int_\Sigma \frac{1}{2}\mathcal{F}\wedge\star_4\mathcal{F}=Q^{D3}_\Sigma+Q^{\overline{D3}}_\Sigma,\label{eqn:induced tension}
\end{equation}
which perturbs the warp factor $h$ in the metric \eqref{eqn:metric ansatz} significantly, does not vanish. As a result, the warped volume of a divisor in the internal space, and so too the Pfaffian, receive corrections depending on \eqref{eqn:induced tension}, and this leads to new inflaton-dependence of the moduli potential.

This effect is not necessarily the most stringent restriction on fluxbrane inflation.  Examining a toroidal orientifold $T^4\times T^2/\Bbb{Z}_2$ for simplicity, \eqref{eqn:induced tension} can be rewritten as
\begin{equation}
Q^{D3}_\Sigma+Q^{\overline{D3}}_\Sigma=2\frac{\mu_7}{\mu_3}\frac{\left( \int_\Sigma J\wedge\mathcal{F}\right)^2}{\frac{1}{2}\int_\Sigma J\wedge J}\,.
\end{equation}
The quantity on the right-hand side is constrained \cite{Hebecker:2012aw} by upper bounds on the cosmic string tension \cite{Urrestilla:2011gr}, which put an upper bound on the D-term potential, and so on the scale of inflation.
The resulting bound is
\begin{equation}\label{eqn:induced tension bound}
Q_\Sigma^{D3}+Q_\Sigma^{\overline{D3}} \lesssim 10^{-1}\,.
\end{equation}
Thus, fluxbrane inflation scenarios whose D-term potential is small enough to avoid upper limits on cosmic strings involve the accumulation of a relatively small D3-brane dipole, and backreaction is not a severe problem.  However, for any variations of fluxbrane inflation that evade cosmic string limits through a mechanism other than reducing the overall scale of inflation, and in which $Q_\Sigma^{D3}+Q_\Sigma^{\overline{D3}}$ becomes significant, a detailed study of backreaction would be important.

\section{Conclusions}  \label{sec:conclusions}

Axion monodromy inflation proceeds via the progressive discharge of $N>1$ units of a quantized charge.  The stress-energy of this monodromy charge sources curvature in the noncompact spacetime, leading to accelerated expansion, but also necessarily sources curvature in the internal six dimensions.  The backreaction effects of monodromy charge on the internal solution are known to be important in the NS5-brane axion monodromy scenario of \cite{MSW}, and were extensively studied in that context in \cite{MSW,FMPWX,MSSSW}, but have not been examined at a comparable level in other models.

In this work we computed the backreaction of monodromy charge in Higgs-otic inflation, an axion monodromy scenario in which inflation is driven by the motion of a D7-brane that becomes magnetized as it travels through a background of three-form flux.  Such a magnetized D7-brane is a localized source in the supergravity equations of motion, and its position and degree of magnetization affect the solution in the internal space.
In \S\ref{sec:backreaction} we obtained the resulting solution, to first order in the perturbation due to the D7-brane, in the case of a toroidal orientifold compactification.
We found that nonperturbative superpotential terms from Euclidean D3-branes or from gaugino condensation depend on the position of the magnetized D7-brane, cf.~\eqref{eqn:the one loop pfaffian} and \eqref{eqn:the one loop pfaffian explicit}.  Thus, the moduli potential depends on the inflaton vev, via the backreaction of induced D3-brane charge on the supergravity solution in the internal space.

Our result echoes the situation in D3-brane inflation, where the position of a mobile D3-brane appears in a Pfaffian factor of the nonperturbative superpotential \cite{Ganor:1996pe,Kachru:2003sx,Berg:2004ek,Baumann:2006th}, and leads to inflaton-dependence of the moduli potential.  Here, however, the D3-brane charge in question is dissolved as flux in a mobile D7-brane; the amount of induced charge changes as the D7-brane moves; and both D3-brane and anti-D3-brane charges contribute.  After a somewhat intricate calculation, our final result is the simple expression~\eqref{eqn:the one loop pfaffian}, in which D3-brane charge and anti-D3-brane charge on the D7-brane, and D(-1)-brane charge on the Euclidean D3-brane, enter on precisely equal footing.

The methods used here apply with little modification to any scenario of axion monodromy in which the inflaton is the position of a mobile brane, and in which there are important nonperturbative contributions to the moduli potential.  We expect comparably strong backreaction effects in such models. However, our results do not constrain axion monodromy scenarios stabilized by purely perturbative effects, nor do they apply to scenarios such as \cite{McAllister:2014mpa} in which the monodromy charge is dispersed in the six-dimensional bulk rather than localized on a brane.

Our findings present an obstacle to achieving D7-brane axion monodromy inflation in a stabilized string compactification, but in our view they do not give such models a uniquely problematic status.  Instead, our results show that F-term axion monodromy constructions such as Higgs-otic inflation face the same challenges as the NS5-brane models of \cite{MSW}, and manifest in these models the well-known couplings of moving branes to nonperturbative superpotential terms that plague D3-brane inflation scenarios \cite{Kachru:2003sx,Berg:2004ek,McAllister:2005mq,Baumann:2006th,Baumann:2014nda}.  In short, the backreaction problem that we find in D7-brane axion monodromy inflation
has causes and severity that precisely match what we would expect based on studies of kindred models.

In view of our findings, it would be worthwhile to search for a mechanism that can alleviate the backreaction of monodromy charge in D7-brane monodromy models.  More generally, exhibiting an explicit and arbitrarily well-controlled solution of string theory that supports large-field inflation remains an important problem.

\section*{Acknowledgments}
We thank Eva Silverstein, John Stout, and Irene Valenzuela for helpful discussions, and we thank Irene Valenzuela for comments on a draft.  The work of M.K.~and L.M.~was supported in part by NSF grant PHY-1719877.

\begin{appendix}
\section{Conventions for Differential Forms}\label{app:convention}

The orientations of D7-branes, and the self-duality properties of two-form fluxes on them, are crucial in D7-brane monodromy models.  We therefore devote this Appendix to laying out our conventions for differential forms, orientation, and the Hodge star operator.

Consider an orientable Riemannian manifold $X$ of real dimension $2d$.
Given an orientation on $X$, and equipped with the natural inner product $ \langle\_,\_\rangle$ such that
\begin{equation}
\langle\_,\_\rangle : \Lambda^p TX^* \times \Lambda^p TX^*\rightarrow \Bbb{C},~ (\omega,\nu)\mapsto \langle\omega,\nu\rangle,
\end{equation}
we define the Hodge star map for differential $p$-forms$\ \omega$\ and$\ \nu$\ as a map
\begin{equation}\label{realstar}
\star_{2d} : \Lambda^p TX^*\rightarrow \Lambda^{2d-p}TX^*,
\end{equation}
such that
\begin{equation}
\omega\wedge\star_{2d}\nu=\langle \omega,\nu\rangle \text{Vol}_{2d},
\end{equation}
where $\text{Vol}_{2d}$ is the volume form of $X$ with the given orientation.
There is a natural generalization of the Hodge star \eqref{realstar} in the case
that $X$ is a complex manifold of complex dimension $d$.  Taking $\omega,$ $\nu$ to be elements of $ \Lambda^p TX^*\wedge \Lambda^q \overline{TX}^*$, the Hodge star map is a linear map
\begin{equation}\label{cxstar}
\star_d:\Lambda^pTX^*\wedge\Lambda^q\overline{TX}^*\rightarrow \Lambda^{d-q}TX^*\wedge \Lambda^{d-p}\overline{TX}^*,
\end{equation}
such that
\begin{equation}
\omega\wedge\star_d\overline{\nu}=\langle \omega,\overline{\nu}\rangle \text{Vol}_d.
\end{equation}
The definitions \eqref{realstar}, \eqref{cxstar} agree on real differential forms and there is no ambiguity regarding the definition of the Levi-Civita symbol.

Under a change of the orientation, the volume form changes sign, and hence so do the eigenvalues of the Hodge star.  Taking $d=3$, a fixed three-form flux that is ISD for one orientation of $X$ is IASD for the opposite orientation.  Likewise, taking $X$ to be a divisor of a threefold ($d=2$), a fixed two-form flux that is SD in one orientation is ASD for the opposite orientation.
Thus, to give a correct description of D-branes in a flux compactification on a threefold $X$, we must specify a set of internally consistent conventions for the orientation of $X$, the orientation of divisors $D\subset X$, and the definitions of $\star_6$ and $\star_4$.  We will now work out the relations among these definitions.

We begin with a canonical choice of orientation, and show which other choices are logically possible.
For $X$ a K\"ahler manifold, we write the K\"ahler form $J$ in
local coordinates as
\begin{equation}\label{kformis}
J=i g_{a\bar{b}}dz^a\wedge d\bar{z}^{\bar{b}}.
\end{equation}
It is natural to define the volume form, and thus the orientation of the manifold, as
\begin{equation}\label{volformis}
\text{Vol}_d=\frac{1}{d!}J^d,
\end{equation}
where in local coordinates with diagonalized metric the volume form is written as
\begin{equation}
\text{Vol}_d= i^d\det(g_{a\bar{b}})  dz^1\wedge d\bar{z}^1\cdots dz^d\wedge d\bar{z}^d.
\end{equation}
We then call the orientation constructed above the \emph{canonical orientation}. For example, the canonical orientation of the volume form on a manifold $X$ with $d=3$ is
\begin{equation}
-i dz^1\wedge d\bar{z}^1\wedge dz^2\wedge d\bar{z}^2\wedge dz^3 \wedge d\bar{z}^3.
\end{equation}
Correspondingly, if $D \subset X$ is a submanifold of complex dimension two, and is dual to a curve of positive volume, then
the orientation on $D$ is
\begin{equation}
-dz^1\wedge d\bar{z}^1\wedge dz^2\wedge d\bar{z}^2.
\end{equation}
From the definition of the Hodge star map, an SD real two-form $\mathcal{S}$ and an ASD real two-form $\mathcal{A}$ satisfy the following relations:
\begin{align}
& \mathcal{S}\wedge \mathcal{S}=\mathcal{S}\wedge \star \mathcal{S}= \langle \mathcal{S},\mathcal{S}\rangle \text{Vol},\\
& \mathcal{S}\wedge \mathcal{A}=0,\\
&\mathcal{A}\wedge \mathcal{A}=-\mathcal{A}\wedge \star\mathcal{A}=-\langle \mathcal{A},\mathcal{A}\rangle\text{Vol}.\label{eqn:antiself}
\end{align}
The  K\"ahler form in a manifold with $d=2$ is SD in the canonical orientation, as $\langle J,J\rangle =2$.

Taking the definition of the Hodge star to be \eqref{cxstar}, one finds that a flux of Hodge type $(2,1)_{\text{primitive}}+(0,3)$ is ISD --- a relation that is ubiquitous in the literature on flux compactifications --- and similarly a flux of Hodge type $(2,0)+(0,2)$  on $D \subset X$ is SD.  These results confirm that our conventions \eqref{kformis},\eqref{volformis}, and \eqref{cxstar} for orientation and for the Hodge star in K\"ahler manifolds are compatible with the literature.

For completeness let us nevertheless explore other possible choices of consistent conventions: see Table \ref{table:convention}.
We will impose a few requirements, which imply conditions on the numbers $a,b \in \{\pm 1\}$ appearing in Table \ref{table:convention}.
The first requirement is that the integral of the volume form over a positively-oriented manifold must be positive.
We will also require that forms of Hodge type $(2,1)_{\text{primitive}}+(0,3)$ are ISD rather than IASD, which implies $ab=1$.
A final requirement is that the bulk Chern-Simons coupling $\propto \frac{1}{i}\int G\wedge \bar{G}$ for forms of type $(2,1)_{\text{primitive}}+(0,3)$ should correspond to positive D3-brane charge whose sign is $b$.
Given these physics inputs, the following describe self-consistent conventions. First, spacetime-filling Dp-brane actions are of the form
\begin{equation}
-\mu_p \int \im\tau\text{Vol}_{p+1}+b\mu_p\int C_{p+1}.
\end{equation}
The Bianchi identity for the RR 4-form field is
\begin{equation}
d\tilde{F}_5=H\wedge F-b\rho^{D3},
\end{equation}
where $ \rho^{D3}$ is the D3-brane charge density. If $G$ is ISD, then
\begin{equation}
H\wedge F= -\frac{G\wedge\bar{G}}{2i\im\tau}= -b\frac{|G|^2\text{Vol}}{2\im\tau}.
\end{equation}
In an ISD background, the following quantity vanishes:
\begin{equation}
\Phi_{-1 \cdot b}=h^{-1}-b\alpha,
\end{equation}
where $ h$ is the warp factor and$\ C_4=\alpha dx^0\wedge dx^1\wedge dx^2\wedge dx^3.$
In this paper, we have taken $a=b=1$.

For \emph{any} choice of $a$ and $b$, the orientation on an effective divisor $D$ is $\frac{1}{2} J\wedge J$, and a form of type $(2,0)+(0,2)$ on $D$ is self-dual on $D$, and so induces D3-brane charge, rather than anti-D3-brane charge, on a D7-brane wrapping $D$.

\begin{table}[!]
\centering

\begin{tabular}{|c|c|c|}
\hline
 & + & - \\ \hline
Choice of Hodge star ($a$) &  $\omega \wedge \star \bar{\nu}=\langle \omega,\bar{\nu}\rangle \text{Vol}$ &  $\star\omega \wedge  \bar{\nu}=\langle \omega,\bar{\nu}\rangle \text{Vol}$ \\ \hline
Choice of  K\"ahler form ($b$) & $ig_{z_i\bar{z}_j}dz^i\wedge d\bar{z}^j$   &  $-ig_{z_i\bar{z}_j}dz^i\wedge d\bar{z}^j$ \\ \hline
\end{tabular}
\caption{Possible conventions. The first column denotes the quantity whose definition can be chosen.  The variables $a$ and $b$ in parentheses equal $+1$ if the choice corresponds to the second column and $-1$ if the choice corresponds to the third column.
We have taken $a=b=1$ throughout this work.}
\label{table:convention}
\end{table}

\section{Green's Function on a Toroidal Orientifold}\label{app:green}
In this section, we provide the Green's function on a simple toroidal orientifold.
The Green's function on $T^2$ is very well known --- see e.g.~\cite{Polchinski:1998rq}.
Here we will provide modular invariant Green's functions on orbifolds and orientifolds of $T^2$ and $T^6$.

Finding a Green's function on a compact manifold of real dimension greater than two by the method of images can be challenging, as the sum diverges in general. In order to deal with this divergence, we regulate the Green's function on a torus. Given this regularized Green's function, we extend it to a Green's function on an orbifold and an orientifold.

We begin with a $T^6$ obtained by identification of the opposite faces of the six-cube of side length $L$.
We then define a toroidal Green's function to be a function that satisfies
\begin{equation}
\nabla^2 G_{(6)}(x;x')=\delta_{(6)}(x-x')-\frac{1}{\int_{T^6 }\text{Vol}_6}.
\end{equation}
The Green's function for the torus is then written as
\begin{equation}
G_{(6)}(x;0)=-\sum_{n\in\Bbb{Z}^6}(1-\delta_{n,0})\frac{e^{2\pi i \vec{n}\cdot \vec{x}/L}}{4\pi^2 n^2 L^4}.
\end{equation}
As we anticipated above, this sum diverges. We follow a prescription given in \cite{Shandera:2003gx} to regularize the Green's function:
\begin{align}
G_{(6)}(x;0)=&L^{-4}\int_0^{\infty}\sum_{n\in\Bbb{Z}^6}(1-\delta_{n,0}) e^{2\pi i \vec{n}\cdot \vec{x}/L-4\pi^2 n^2 s}ds\label{eqn:6d green}\\
=&L^{-4}\int_0^\infty\left(1- \prod_{j=1}^6\sum_{n\in\Bbb{Z}^6}e^{2\pi i n_j x_j/L -4\pi^2 n_j^2 s} \right)ds\\
=&L^{-4}\int_0^\infty \left( 1-\prod_{j=1}^6\vartheta_3 \left(\left.\frac{ x_j}{L}\right|4\pi i s\right)\right)ds.
\end{align}
We used the identity
\begin{equation}
\vartheta_3\left(\nu|\tau\right)=\sum_n e^{2\pi i (\nu n+\tau n^2/2)}
\end{equation}
for the last equality.

In order to obtain lower-dimensional toroidal Green's functions, we dimensionally reduce the six-dimensional Green's function \eqref{eqn:6d green}. It is then clear that the Green's function satisfies the identity
\begin{equation}
\int d^d x G_{(6)}(x;x')=G_{(6-d)}(x;x').
\end{equation}
We choose $ G_{(0)}(x;x')=0.$  We expect that $G_{(2)}(z;z')$ would correspond to the well known toroidal Green's function
\begin{equation}
G^{(2)}(z;z')=\frac{1}{2\pi}\log \left| \vartheta_{1} \left(\left.\frac{z-z'}{L}\right|\tau \right)\right|-\frac{(\im( z-z'))^2}{2L^2\im \tau}+C(\tau),\label{eqn:Green's function on a 2d torus}
\end{equation}
where $\tau$ is the complex structure modulus, and $C(\tau)$ is a function of $\tau$ \cite{lin2010elliptic} that must obey
\begin{align}
&C(\tau+1)=C(\tau),\\
&C(-1/\tau)=C(\tau)-\frac{1}{4\pi}\log\left| \tau\right|,
\end{align}
in order for the Green's function to be invariant under modular transformations.
These modular transformation properties suggest that $C(\tau)$ is given by
\begin{equation}
C(\tau)=-\frac{1}{2\pi}\log\left|\eta(\tau)\right|+C_0,\label{eqn:modularity}
\end{equation}
where $\eta(\tau)$ is the Dedekind eta function and $C_0$ is undetermined constant. We determined $C_0=0$ numerically by demanding that the integral of the Green's function \eqref{eqn:Green's function on a 2d torus} over the torus vanishes.

Given the toroidal Green's function (\ref{eqn:6d green}), it is natural to extend it to the Green's function defined on a toroidal orbifold or a toroidal orientifold. Let us work with an example for simplicity. For a finite group $\Bbb{Z}_N,$ let there be a group action $\theta$ on a complex coordinate $z.$ Then we denote a Green's function defined on the toroidal orbifold/orientifold $T^6/\Bbb{Z}_N$ as
\begin{equation}
G_{T^6/\Bbb{Z}_N}(z;z')=\sum_i^N G_{(6)}(z;\theta^i z').\label{eqn:Green's function on toroidal orientifolds}
\end{equation}
Similarly, a Green's function on $T^2/\Bbb{Z}_N$ is determined as
\begin{equation}
G_{T^2/\Bbb{Z}_N}(z;z')=\sum_i^N G_{(2)}(z;\theta^i z').\label{eqn:2d Green's function on toroidal orientifolds}
\end{equation}
Here $z$ and $z'$ are understood to be in the fundamental domain.  We frequently omit the subscript $T^2/\Bbb{Z}_N$.

Finally, we will make use of the identity
\begin{align}
\int d^dx' \nabla G_{(d)}(x;x') \cdot \nabla G_{(d)}(x';x_0)=&-\int d^d x' G_{(d)}(x;x') \nabla^2 G_{(d)}(x';x_0)\\
=&-G_{(d)}(x;x_0).
\end{align}

\end{appendix}
\bibliographystyle{JHEP}
\bibliography{refs}
\end{document}